\documentstyle[12pt]{article}

\author
{S. N. Dorogovtsev \thanks{e-mail: sn@dor.ioffe.rssi.ru}\\
{\it A. F. Ioffe Physical and Technical Institute,} \\
{\it 194021, St. Petersburg, Russia}}
\title{Avalanche Mixing of Granular Solids
}

\date{}

\begin{document}

\maketitle

\noindent
PACS. 64.75 -- Solubility, segregation and mixing. \\
PACS. 05.40 -- Fluctuation phenomena, random processes
and Brownian motion. \\
PACS. 46.10 -- Mechanics of discrete systems.

\renewcommand{\abstractname}{}
\begin{abstract}
\noindent
{\bf Abstract.} --
Mixing of two fractions of a granular material in a slowly rotating
two-dimensional drum is considered.
The rotation is around the axis of the upright drum.
The drum is filled partially, and mixing occurs only at a free surface
of the material.
We propose a simple theory of the mixing process
which describes a real experiment surprisingly well. The dependence
of the mixing time on the drum filling is calculated.
The mixing time is infinite in the case of the half-filled drum.
We describe singular behaviour of the mixing near this
critical point.
\end{abstract}

A beautiful experiment on so called avalanche
mixing of granular solids in a slowly
rotated two-dimensional drum was presented in the paper
of Metcalfe {\it et al}
\cite{metcal}. The patterns observed were so striking
that one of the figures made the cover of the March (1995) issue of
{\it Nature}. In the present communication, we shall propose a simple
theory which provides a surprisingly good description of this experiment.
We shall demonstrate that kinetics of avalanche mixing
can be understood using a geometrical approach without appealing to ideas
of self-organized criticality \cite{bak1}.
(One should note that the
flow of granular solids in rotated drums is usually used to study
self-organized criticality
\cite{buch,po,cant,koh,ris,raj,baum2,zik,hil,fret,chris},
{\it etc.})
In fact, we shall find the main quantity appearing in the problem ---
the mixing time --- for all values of the drum filling.

Two types of small granules --- black and white --- are poured into a
two-dimensional upright drum.
Except of their colours, the granules of the different fractions are
indistinguishable. The drum is filled partially, so there is a free space
at the top.
The specific initial distribution of pure fractions will be
insignificant for our aims.
(In the experiment \cite{metcal},
at the first moment, the black fraction is above the
white fraction --- see fig. 1, {\it a}.)
The drum starts to rotate adiabatically slowly
around its axis in a counter-clockwise direction, for
definiteness. Therefore, the complete angle of rotation plays the role
of time. The answer will not depend on the radius of the drum, which may
be assumed equal to 1.
It is convenient to characterize the
relative volume of the unfilled space by the angle $\theta$
(see fig. 1, {\it b}).
We describe macroscopically the state of each point of the material
by the quantity
$\rho$ --- the concentration of the black fraction at a given location
(for a pure black fraction we have $\rho=1$ and for a pure white fraction
$\rho=0$). Thus, we shall study
the evolution of the space distribution $\rho({\bf x},t)$.

Let us assume that the granules cannot intersperse and slip mutually
as long as they are in the bulk of the material. Only the
granules at the free surface layer flow
and undergoing mixing through the action of successive avalanches.
Such mixing is called avalanche mixing \cite{metcal}. In a real experiment,
the free surface is flat, and when the drum rotates slowly, the surface
is tilted at the angle of repose (strictly speaking, the angle fluctuates
slightly with time \cite {metcal}, but these variations are small, and we
shall neglect them). The angle of repose does not appear in our answers,
so we set it equal to zero in fig. 1, {\it b}.

The described simple configuration
demonstrates clearly the essence of avalanche mixing.
One can see immediately that if the drum is more than half full,
the central
part of the material --- it is surrounded by a dashed line in fig. 1 ---
rotates with the drum without any mixing, and mixing occurs only
inside the ring $\cos\theta<r<1$. If the drum is less than half full, the
whole of the material will mix.

Avalanche mixing proceeds in a fastest possible way, if the
material is completely mixed after
interspersion from the right half of the
free surface to the left half (see fig. 1, {\it a}).
For given filling of the drum,
the mixing time for any material undergoing avalanche mixing cannot be
shorter than the mixing time obtained under this assumption.
In this case,
there may be inhomogeneous distribution of the fractions at the
right half of the free surface, but the material becomes to be
mixed homogeneously at the left half.
After some time $\delta t< 2\pi$ the material from the left half of the free
surface turns as a unit to a definite tangent like that
denoted as $CD$ in fig. 1, {\it b}.
Therefore, at sufficiently long time $t>2(\pi-\theta)$, the
concentration $\rho$ is constant at all
points of each separate tangent to the dashed-line circle
(see, {\it e.g.}, the tangent $CD$ in fig.1, {\it b}).
On the other hand, the values
of $\rho$ may be different for the different tangents. Now it is possible to
describe exhaustively the state of our system at the time $t>2(\pi-\theta)$
by the quantity
$\rho(\varphi,t)$, where $\varphi$ is the angle of the corresponding
radius-vector with the normal to the free surface.
Thus, in fact,
the problem becomes to be a pure geometrical one, and there is no
need to take into account all subtleties associated with the
structure and shape of the granules.
We shall consider kinetics of avalanche mixing in this
fastest possible (for avalanche mixing)
regime and compare the results with the experimental data \cite{metcal}.
As it will be shown, the regime under consideration is very close to a real
one.

Let us derive equations for the mixing dynamics. First we consider
the case in which the drum is more than half full ($\theta<\pi/2$) and
$t>2(\pi-\theta)$. (Avalanche mixing at shorter times is discussed in our
papers \cite{dor}.) Let us turn the drum by a small angle $dt$
at the moment $t$.
Then the following amount of the black fraction will flow
from right to left (from sector $A$ to sector $B$ in
fig. 1, {\it a}):
\begin{eqnarray}
\label{e002}dt\,\frac{\sin^2 \theta}2\,\rho(0,t)=
dt\int_0^{\sin\theta} dr r \rho_{right}(r,t)=
\ \ \ \ \ \ \ \ \ \ \ \ \ \ \ \ \ \ \\[1ex]
\nonumber
=dt\! \int_0^{2\theta}d\zeta \frac{d r(\zeta)}{d\zeta} r(\zeta)
\rho_{right}(r(\zeta),t)=
dt\,\frac{\cos^2\theta}2 \int_0^{2\theta}d\zeta
\frac{d \tan^2(\zeta/2)}{d\zeta} \rho(2\pi-\zeta,t)   .
\end{eqnarray}

\noindent
The first integration $\int_0^{\sin\theta}dr$ is from the free surface
centre to the drum wall.
$r(\zeta)$ is the coordinate of the intersection of the tangent fixed by
the angle $\varphi=2\pi-\zeta$ with the free surface (see the segment
$ED$ in fig. 1, {\it b}). Note that
$r(\zeta)=\cos\theta \tan(\zeta/2)$, so $r(2\theta)=\sin\theta$.
Thus, the concentration of the black fraction at the different points
of the right half of the free surface equals
$\rho_{right}(r(\zeta),t)=\rho(\zeta,t)$. One can see that
$\rho(\varphi,t)=\rho(0,t-\varphi)$ when $t\geq\varphi\geq 0$,
because granules cannot mix inside the material. Thus, from
eq. (\ref{e002}), we obtain the following equation for $\rho(0,t)$:
\begin{equation}
\label{e003}
\rho(0,t)=
\cot^2\theta \int_0^{2\theta}d\zeta
\rho(0,t-2\pi+\zeta) \frac{\sin(\zeta/2)}{\cos^3(\zeta/2)}  \ .
\end{equation}

The case in which the drum is less than half full may be considered
in the same way.
After introducing the angle $\vartheta\equiv\pi-\theta$ we obtain the
equation:
\begin{equation}
\label{e004}
\rho(0,t)=
\cot^2\vartheta \int_0^{2\vartheta}d\zeta
\rho(0,t-\zeta) \frac{\sin(\zeta/2)}{\cos^3(\zeta/2)}  \ .
\end{equation}

\noindent
Thus, a complicate dynamical problem is reduced to studying of
dynamics of the linear zero-dimensional system! Note that eqs. (\ref{e003})
and (\ref{e004}) are not interconverted after the formal change
of variables $\theta=\pi-\vartheta$ because of the non-analyticity
at the half-filling point $\theta=\vartheta=\pi/2$.
Indeed, if the drum is exactly half full, it follows immediately from eq.
(\ref{e003}) or (\ref{e004}) that $\rho(0,t)=\rho(0,t-\pi)$. Therefore the
material will never be mixed to a uniform state, and the mixing time turns
to be infinite.

Let us study asymptotic behaviour of the solutions of eqs. (\ref{e003}) and
(\ref{e004}). It is sufficiently to substitute $\rho(0,t)=e^{zt}$ into
them. If the drum is more than half full, one gets from eq. (\ref{e003}):
\begin{equation}
\label{e005}
e^{2\pi z}=2\cot^2\theta \int_0^{\theta}d\xi
e^{2z\xi} \frac{\sin\xi}{\cos^3\xi}  \ .
\end{equation}

\noindent
If the drum is less than half full, it follows from eq. (\ref{e004}) that
\begin{equation}
\label{e006}
1=2\cot^2\vartheta \int_0^{\vartheta}d\xi
e^{-2z\xi} \frac{\sin\xi}{\cos^3\xi}  \ .
\end{equation}

We have derived two transcendental equations for $z$. To obtain the
characteristic time of mixing one should find their nearest to zero
roots. These roots are expressed by the following relation:
\begin{equation}
\label{e007}
z=-\tau^{-1}\pm i 2\pi/T \ ,
\end{equation}

\noindent
where $\tau$ and $T$ are the mixing time and the oscillation period,
correspondingly. Therefore, at long times, the concentration of the black
fraction at the left half of the free surface evolves in the following
way:
\begin{equation}
\label{e001}
\rho(0,t)=\rho_\infty +
C_0\exp\{-t/\tau\}\cos\left\{2\pi t/T+\varphi_0\right\} \ .
\end{equation}

\noindent
Here $\rho_\infty$ is the concentration of the black fraction in the
homogeneously mixed material. $\rho_\infty$, $C_0$, and $\varphi_0$ are
constants depending on the initial conditions, that is on the amount and
distribution of the pure fractions at the initial moment.
One can see that $\tau$ and $T$ depend only on one parameter, that is on
the drum filling.
Let us find the roots of eqs. (\ref{e005}) and (\ref{e006}).

Though the solutions of the transcendental equations
are not expressed analytically, they may be found
with any desired
precision. The integrals in eqs. (\ref{e005}) and (\ref{e006}) may be
expressed in terms of special functions. Since these functions are rather
exotic, we shall calculate directly the integrals in eqs. (\ref{e005})
and (\ref{e006}). For example, one may integrate them
by parts and then iterate the following relations:
\begin{equation}
\label{e008}
z=\frac1{2\pi}\log{\left\{e^{2\theta z}-2z\cot^2\theta
\int_0^\theta d\xi e^{2z\xi} \tan^2\xi\right\}}
\end{equation}

\noindent
if the drum is more than half full $(\theta<\pi/2)$ and
\begin{equation}
\label{e009}
z=\frac1{2\vartheta}\log{\left\{1+2z \cot^2\vartheta e^{2z\vartheta}
\int_0^\vartheta d\xi e^{-2z\xi} \tan^2\xi \right\}}
\end{equation}

\noindent
if the drum is less than half full $(\vartheta<\pi/2)$.

The iterations converge very fast. Thus, using eq. (\ref{e007}), we
obtain the dependence of the inverse mixing time $\tau^{-1}$ on the value
of the drum filling (see fig. 2), that is our main result.
We express $\tau^{-1}$ through $v=\vartheta-\sin{2\vartheta}/2$,
 {\it i.e.} the fractional volume of the drum which is occupied by the
material. For comparison, in fig. 2, we also show
the experimental points taken from the paper \cite{metcal}.

The obtained $\tau(v)$ is the shortest possible time of avalanche mixing
at the given filling $v$. Thus, the curve $\tau^{-1}(v)$ for any
granular material must be under the extreme curve in fig. 2.
One sees that though the experimental points from the paper \cite{metcal}
are near the extreme dependence, in fact, they are under it.
We find that the experiment \cite{metcal} is very close to the
regime of the fastest avalanche mixing. Therefore,
it is described by our theory,
which does not include any parameters of the mixing granular material.

One may introduce the mixing rate $v/\tau$
\cite{metcal}. In our theory, its maximum value equals
$\max(v/\tau)=0.0683\ldots$ at $v=0.177\ldots$
These values are also close to the experimental ones:
$\max(v/\tau)=0.056\pm0.006$ at $v=0.17\pm 0.015$ \cite{metcal}.

The analytical results for $\tau^{-1}$ and $T$ for the limit cases of
low and high filling are known already \cite{dor}.
For low  filling $\vartheta \ll 1$ we have obtained
$\tau^{-1}\cong1.392/2\vartheta$ and $T\cong4\pi\vartheta/7.553$ and for the
almost full drum $(\theta \ll 1)$ ---
$\tau^{-1}\cong \frac19\frac1{2\pi}\theta^2$ and $T\cong2\pi-4\theta/3$.

Let us discuss the behaviour of $\tau$ and $T$ near the half filling
point. If the drum is exactly half full, the
concentration of the black fraction is a periodic function with the period
$\pi$, {\it i.e.} the half period of the rotation of the drum, and
the material will never be mixed. Thus, let us write the root sought in the
form $z=2i+s$ where $s$ is a small addition. First we consider
the case in which the drum is more than half full. Substituting
this relation into eq. (\ref{e003}) we obtain
\begin{equation}
\label{e010}
e^{2\pi s}=2\cot^2\theta\! \int_0^{\theta}d\xi e^{2s\xi}
\left\{4\sin2\xi+\tan^3\xi-7\tan\xi-4 i(\cos2\xi+\tan^2\xi-1) \right\}.
\end{equation}

\noindent
Now we introduce a small deviation $\delta$ from the half filling:
$\theta=\pi/2-\delta$, and then one may expand $e^{-2s\xi}$ in small $s\xi$.
Using the integration by parts it is easy to select the singular at
$\delta\to 0$ terms. The rest (finite at $\delta\to 0$) integrals are
also evaluated without trouble.
Then one should compare the coefficients of the
terms having equal powers $\delta$.

After the straightforward calculations one gets
\begin{equation}
\label{e011}
z=2i-\frac{8\delta}\pi i -
\frac{16\delta^2}\pi \left[\log(
1/|\delta|)-5/2\right]+\ldots
\end{equation}

\noindent
Taking into account eq. (\ref{e007}), we find from eq. (\ref{e011})
that at sufficiently long times $(t\gg \tau)$ mixing evolves according
to the law (\ref{e001})
in which the characteristic mixing time $\tau$ and the oscillation period
$T$ are expressed in terms of $\delta=\pi/2-\theta$:
\begin{equation}
\label{e012}
\tau^{-1}=\frac{16\delta^2}\pi \left[\log(
\frac1{|\delta|})-\frac52\right]
\end{equation}

\noindent
and
\begin{equation}
\label{e013}
T=\pi+4\delta \ .
\end{equation}

\noindent
$\tau$ and $T$ may also be expressed in terms of $v-1/2\cong 2\delta/\pi$.

Similar calculations may be done for less than half filling.
The results are the same as (\ref{e012}) and
(\ref{e013}) (now $\delta=\vartheta-\pi/2<0$).
Note that such a coincidence is unusual for critical phenomena \cite{zin}.
Analytical expression (\ref{e012}) for $\tau(v)$ is available only
in the narrow region $|v-1/2|<0.02$.

In our paper we neglect fluctuations of the inclination angle of the free
surface. In fact, this angle fluctuates in the interval between the angle
of repose and the angle of marginal stability. If
we shall take into account these fluctuations, the results will change
slightly. For instance,
in the limit of low filling $v\to 0$ one obtains nonzero $\tau$.

In summary, as we have shown above, in the experiment \cite{metcal},
mixing
proceeds in a nearly fastest possible (for avalanche mixing) way. In
this case, the description of avalanche mixing becomes to be very
simple --- the problem is reduced to studying of the dynamics of a linear
zero-dimensional system. In fact, we have solved it analytically.
One see
that the system under consideration demonstrates unusual critical
behaviour near the half filling point at which the mixing time is infinite.
Note finally, we have described quantitatively the beautiful experiment
without using any characteristics of the granules.

\medskip
The author thanks Yu.A.Firsov, A.V.Goltsev, S.A.Ktitorov,
E.K.Kudinov, A.M.Monakhov, and B.N.Shalaev for many helpful discussions.
This work was supported in part
by the RFBR grant.

\newpage

\newpage
S. N. Dorogovtsev

Avalanche Mixing of Granular Solids \\
\vspace{12mm}

\begin{center}
FIGURE CAPTIONS \\
\end{center}
\vspace{8mm}

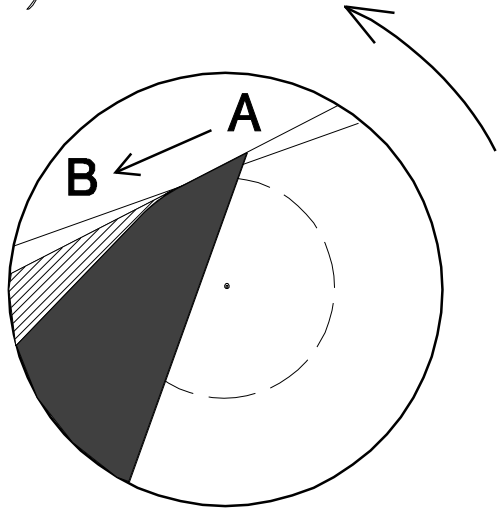
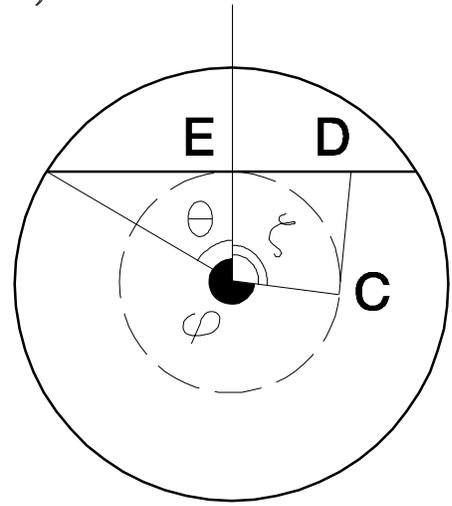
Fig. 1. -- {\it a}) The avalanche mixing scheme.
For an infinitely small turn of the drum, the granules of different
fractions flow from sector $A$ to sector $B$, undergoing mixing.
The free surface of the material is tilted at the angle of repose all the
time. The pure fractions are shown by the black and white colours.
Regions of mixed material are denoted by the grey colour.
The degree of mixing is not shown.
	{\it b}) The arrangement of the material for the case in which
the drum is more than
half full. The free surface is shown to be horizontal, since our
results do not depend on the angle of repose.
The angle $\theta$ characterizes the relative volume of the free space.
The position of the tangent $CD$ is fixed by the angle $\varphi$.
For the fastest avalanche mixing, the concentration of the black fraction
at all points of such a tangent equals
$\rho(\varphi,t)$ at sufficiently long times.
The tangent crosses the free surface if
$\zeta<2\theta$, where $\zeta \equiv 2\pi-\varphi$.

\vspace{8mm}

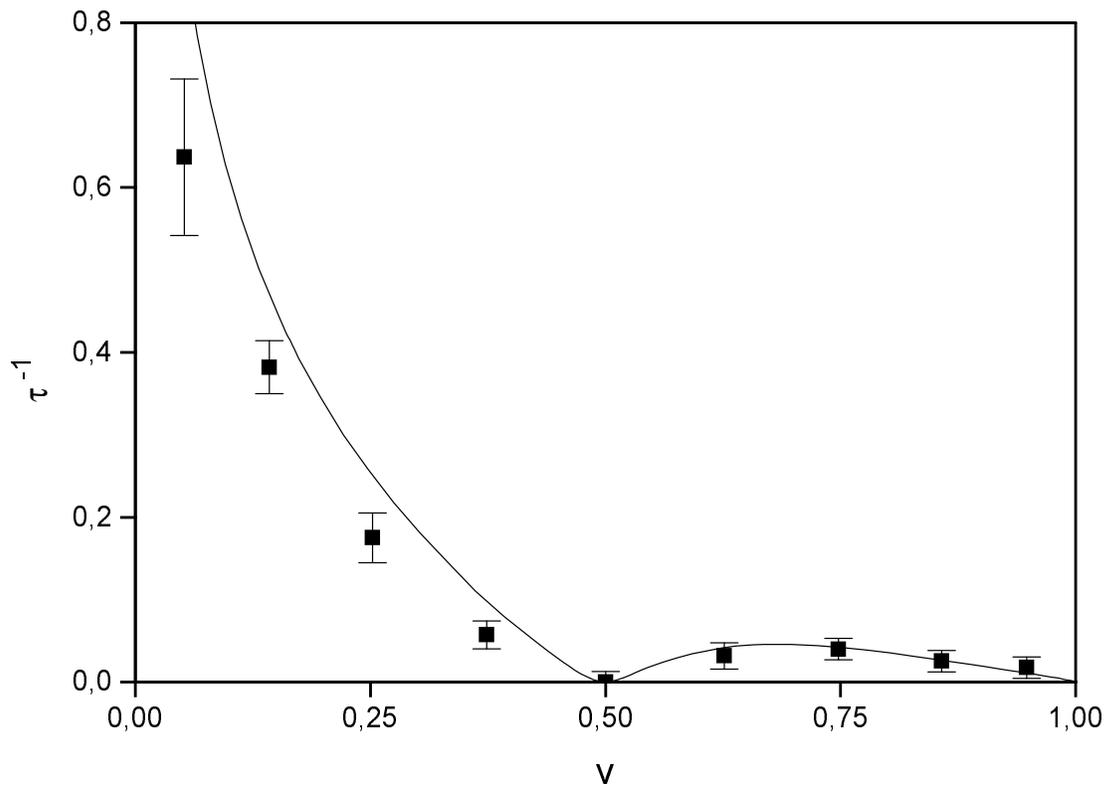
Fig. 2. -- Inverse characteristic time of mixing as a function of
the relative filled volume of the drum. For comparison,
the experimental points from the paper \cite{metcal} are shown.

\end{document}